\documentclass[prd,eqsecnum,apsi]{revtex4}
\usepackage{graphicx}
\usepackage{latexsym}
\usepackage{amsmath}
\usepackage{amssymb}

\newcommand{\beq}{\begin{equation}}
\newcommand{\eeq}{\end{equation}}
\newcommand{\bea}{\begin{eqnarray}}
\newcommand{\eea}{\end{eqnarray}}
\newcommand{\bfig}{\begin{figure}}
\newcommand{\efig}{\end{figure}}

\newcommand{\cosech}{\hbox{cosech}}

\begin{document}
\title{Kaluza-Klein modes of bulk fields in a generalized Randall-Sundrum scenario}
\author{Joydip Mitra \footnote{Electronic address: {\em tpjm@iacs.res.in}}
${}^{}$and Soumitra SenGupta \footnote{Electronic address : {\em tpssg@iacs.res.in}
}${}^{}$}
\affiliation{Department of Theoretical Physics and Centre for
Theoretical Sciences,\\
Indian Association for the Cultivation of Science,\\
Kolkata - 700 032, India}

\begin{abstract} We consider a generalised two brane Randall-Sundrum model with 
non-zero cosmological constant on the visible TeV brane. 
Massive Kaluza-Klein modes for various bulk fields namely graviton, gauge field and  antisymmetric second rank  
Kalb-Ramond field 
in a such generalized Randall-Sundrum scenario are determined.
The masses for the  Kaluza-Klein excitations of different bulk fields are found to depend on the brane cosmological constant
indicating interesting consequences in warped brane particle phenomenology.

\end{abstract}
\maketitle

\section{Introduction}
Theories with extra space-time dimensions are being studied with renewed interest ever-since it was shown that they 
could provide a solution \cite{ADD,cohen}
to the gauge hierarchy problem in connection with the mass of Higgs in the standard model of elementary particles. 
In the warped geometry model proposed by Randall and Sundrum (RS) ~\cite{RS}, one 
considers a single extra dimension compactified on a $S^1/Z_2$ orbifold with two flat 3-branes sitting at the two orbifold fixed points.
The mass scales in the two branes are hierarchically warped from Planck scale to Tev scale. The brane corresponding
to Tev scale is identified as the standard model brane. The brane separation parameter (i.e the modulus)
of such a model can be  stabilized by incorporating a scalar field in the bulk with suitable potential\cite{GW}.
For the localization of the standard model fermion fields on the TeV scale 3-brane, two approaches in general are 
adopted. In one view point the  localization is achieved by a 
bulk scalar field with an appropriate coupling .Alternatively from a string theory angle, the standard
model fermions being open string modes are naturally attached and hence localized to the brane. A closed
string mode like graviton however can propagate inside the bulk. 

In general any bulk field has various Kaluza-Klein ( KK ) mode excitations which are expected to couple to the brane fields leading to 
interesting phenomenology. The roles of such KK modes of different bulk fields on the TeV brane have been explored in    
different works ~\cite{gwpheno,davou,davou1,davou2,davo,agashe} to determine their signatures in the particle phenomenology on the 
standard model brane. Such phenomenological signatures
of course crucially depend on the masses of the Kaluza-Klein modes of these bulk fields. 
It is seen that in the standard RS scenario, the masses  of 
the KK modes of various bulk fields are suppressed by 
the warp factor $e^{-k r_c \pi}$ (where  $r_c$  is the radius of  compactification along the 5-th dimension, k is related 
to the bulk cosmological constant) so that the  
low lying KK modes would be characterized by a scale of the order of TeV which naturally may 
give rise to interesting phenomenology at the TeV scale experiments.
Among various possible fields, scalar, graviton, gauge field and Kalb-Ramond field have drawn special attention
because of their possible presence in the bulk in string-based models.\cite{gsw,lebedev}. The phenomenological 
implications of the KK mode corresponding to these bulk fields have been studied 
extensively in the backdrop of Randall-Sundrum model.\cite{little,bombi,gunion,liam,rizzo1,dominici}

Meanwhile there has been an important generalization of the Randall-Sundrum model. In the original model proposed
by Randall and Sundrum the TeV-scale 3-brane was chosen to be flat i.e with zero cosmological constant. Subsequently this
formalism was generalised to Ricci flat \cite{hawking} as well as  de-Sitter and anti de-sitter 
3-branes\cite{ genrs,christof,kaloper}. It was shown that because of the presence of
non-zero cosmological constant on the 3-brane, the warp factor gets modified and one can have different choices for the value
of $kr$ for different values of the  cosmological constants such that the desired Planck to Tev scale warping can be achieved.
Fermion localization as well as the modulus stabilization of this model have been shown in subsequent works \cite{jm,koley}\\.

In this work we carefully examine the effect of non-zero cosmological constant ( positive or negative) on the KK modes of
the bulk fields which in turn will establish the connection between cosmological constants of our universe at various epoch and the
scenarios where KK modes play an important role \cite{param,muta}. We begin with a brief review of the generalised RS 
model in the next section.
In the subsequent sections we find out the modifications of the KK mode masses of various bulk fields due to the non-zero
cosmological constants on the Tev scale 3-brane.   

\section{Model}

%Generalised RS model
Randall-Sundrum warped braneworld model \cite{RS} has the following features : 
1) The bulk space-time is anti-de-Sitter ( a negative cosmological constant ), the effective 
cosmological constant induced on the TeV/visible brane is zero.\\
2) The brane tension of the standard model/visible brane is negative.\\
3)Without introducing any new scale,other than the Planck scale,  
one can choose the brane separation modulus $r_c$ to have a value $M_P^{-1}$ such that the desired warping can be obtained between the
two branes from Planck scale to TeV scale. This immediately resolves the fine tuning / gauge hierarchy 
problem in connection with the Higgs mass in the standard model.\\  
4)The modulus can be stabilised to the above chosen value 
by introducing scalar in the bulk \cite{GW} without any further fine tuning.\\ 

Question arises that can one generalise such a model with a non-zero cosmological constant on the Tev brane with the possibility
of rendering it with a positive tension without disturbing the main focus of the work namely the resolution of the gauge hierarchy issue.
This was motivated by the facts that the zero cosmological constant of the visible 3-brane is not consistent with the observed 
small value of the cosmological constant of our Universe and negative tension
branes are intrinsically unstable.
Such a generalisation was indeed achieved \cite{genrs}  
and it has been demonstrated  that one can have a more general warp factor which 
includes branes with non-zero cosmological constant \cite{maeda} and in certain cases with positive tension for both  the 
branes. We briefly outline the generalised RS model below.

\section{Generalized RS model}
The  warp factor in such a model is obtained by extremising the action,
\begin{equation}
S = \int d^5x \sqrt{-G} ( M^3 {\cal R} - \Lambda) + \int d^4x
\sqrt{-g_i} {\cal V}_i
\end{equation}
where $\Lambda$ is the bulk cosmological constant, ${\cal R}$ is
the bulk ($5$-dimensional) Ricci scalar and ${\cal V}_i$ is the
tension of the $i^{th}$ brane ($i = hid(vis)$ for the hidden
(visible) brane). It is shown that a  warped geometry results from a constant 
curvature brane space-time, as opposed to a flat 3-brane space-time. The generalized ansatz for the warped metric is given by, 
\begin{equation}
ds^2 = e^{- 2 A(y)} g_{\mu\nu} dx^{\mu} dx^{\nu} +r^2 dy^2 \label{metric}
\end{equation}
where  $r$ corresponds to the modulus associated with the extra dimension   and $\mu , \nu$ stands for brane coordinate 
indices. As in the original RS model, the scalar mass warping is achieved through the warp factor
$e^{-A(k r \pi)}=\frac{m}{m_0}=10^{-n}$ where  $r$ is the compact modulus, $k = \sqrt{- \frac{\Lambda}{12 M^3}}$ $\sim$ Planck 
Mass with the bulk cosmological constant $\Lambda$ is chosen to be negative. `$n$' the 
warp factor index must be set to $16$ to achieve the desired warping 
and the magnitude of the induced cosmological 
constant on the brane in this case is non-vanishing in general and is given by =$10^{-N}$(in Planck units).  
For the induced brane cosmological constant, $\Omega > 0$ and $\Omega < 0$, the brane metric $g_{\mu\nu}$ may
corresponds to some de-sitter or anti de-Sitter space-time for example dS-Schwarzschild and AdS-Schwarzschild 
space-times respectively \cite{karch}.
\subsection{Induced brane cosmological constant $\Omega < 0$}
For AdS bulk i.e.  $\Lambda<0$, considering the regime for which the induced cosmological constant $\Omega$ on the visible brane is 
negative if one redefines $\omega^2 \equiv -\Omega/3k^2 \geq 0$, then the following solution for the warp factor
is obtained :
\begin{equation}
\label{wfads}
e^{-A} = \omega \cosh\left(\ln \frac {\omega} c_1 + ky \right)
\end{equation}
where $c_1 = 1 + \sqrt{1 - \omega^2}$ for the warp factor normalized to unity at $y = 0$. 
One can show that real solution for the warp factor exists if and only if $\omega^2 \leq 10^{-2n}$. 
This leads to an upper bound for the magnitude of the 
cosmological constant as $N_{min}=2 n$. So, for $n=16$ , $\omega^2$ is found to be $10^{-32}$. For $N=N_{min}$, we get a degenerate solution~
$x=n \ln {10}+ \ln{2}$, where $x = kr\pi$. For $N-2n>>1$, the solutions obtained in this case, are
\begin{equation}
x_1=n \ln{10}+\frac{1}{4}10^{-(N-2n)}  ,  x_2=(N-n)\ln{10}+\ln{4} 
\end{equation}
Thus, to have the required Planck to Tev scale hierarchy ( i.e $n=16$ ) one obtains in general two values of $x$ which
correspond to two different values of the brane separation modulus $r$. For example if we take $n = 16$ and $N = 124$,    
\begin{equation}
 k \pi r_1 \simeq 36.84 + 10^{-93}~,~~ k \pi r_2  = 250.07~
\end{equation}

RS value is recovered for $x=n \ln{10}$ and $N=\infty$. Moreover at $x=n \ln{10}+\ln{2} = x_0$ ( say) 
and $N=2 n$, $\omega^2$ reaches it's maximum value. Beyond this the magnitude of $\omega^2$ starts to decrease again.
One can also obtain the tension of the visible brane for the above two solutions. 
When $N=N_{min}=2 n$ i.e $x = x_0$  the  visible brane tension is zero.
For the entire region for which $x$ is less than $x_0$ the visible brane tension is negative while for $x$ greater than
$x_0$ the visible brane tension is positive. . 
\subsection{Induced brane cosmological constant $\Omega>0$}
In this case, the warp factor is given by,
\bea
\label{wfds}
 e^{-A} = \omega \sinh\left(\ln \frac {c_2}{\omega} - ky \right)~,
\eea
where  $\omega^2 \equiv \Omega/3k^2$ and $ c_2 = {1 + \sqrt{1
+ {\omega}^2}}~.$  In this case there is no bound on the value of $\omega^2$, 
and the (positive) cosmological constant on the visible brane can be of arbitrary magnitude. 
Also, there is a single solution of $k r \pi$ whose precise value will depend 
on $\omega^2$ and $n$. The brane tension is negative for the entire range of values of 
the positive cosmological constant. In Fig.(1) it has been shown how $\omega^2$ is related to the 
modulus $kr$. In this plot we have chosen $n = 16$ so that the hierarchy problem can be solved.
%%%%%%%%%%%%%%%%%%%%%%%%%%%%%%%%%%%%%%%%%%%%%%%%%%%%%%%%%%%%%%%%%%%%%%%%%%%%%%%%%%%%%%%%%%%%%
\begin{figure}[h] 
\includegraphics[width=3.330in,height=2.20in]{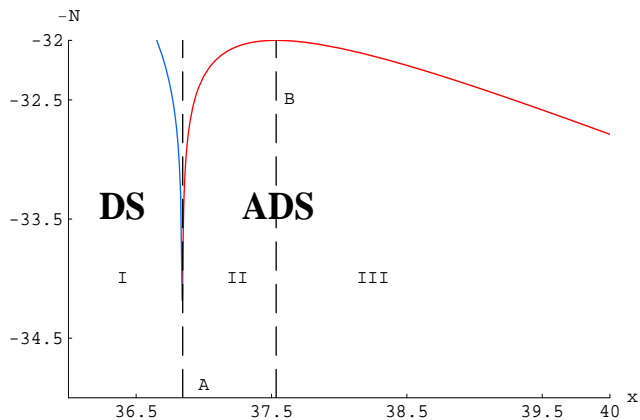}
\caption{Graph of $N$ versus $x = k r \pi =36-40$, for $n=16$ and for both
positive and negative brane cosmological constant. The curve in
region-I corresponds to positive cosmological constant on the brane,
whereas the curve in regions-II \& III represents negative
cosmological constant on the brane. } \label{ADSDS}
\end{figure}
%%%%%%%%%%%%%%%%%%%%%%%%%%%%%%%%%%%%%%%%%%%%%%%%%%%%%%%%%%%%%%%%%%%%%%%%%%%%%%%%%%%%%%%%%%
It is seen that , in both cases, ($\Omega>0$ and  $\Omega<0$) the warp factor depends on brane cosmological constant \cite{genrs}
So we expect that KK modes for different bulk fields in this scenario will depend on the brane cosmological constant ($\Omega$).\\
In this letter, we calculate KK modes and their masses for bulk- gauge field, graviton field and the second rank 2-form anti-symmetric 
Kalb-Ramond field. \\

\section{KK mode for for different bulk fields for anti-de Sitter brane ($\Omega <0$)}
In this case, the warp factor is given by,
\bea 
\label{wfads}
e^{-A} = \omega \cosh\left(\ln \frac {\omega} c_1 + ky \right)
\eea
where $c_1 = 1 + \sqrt{1 - \omega^2}$. 
\subsection{KK Modes for bulk gauge field}
Consider bulk $U(1)$  gauge field  $A_M$(where the index $M$ runs over 5 dimensions).
It's components $A_{\mu}$ (where ${\mu}$ runs over four dimensions) is $Z_2$ even and $A_4$ is $Z_2$ odd 
with respect to extra dimension $y$. $A_4$ therefore does not have any zero mode in the four- dimensional theory  \cite{gross,davou}).
To start with, we have five-dimensional action $S_A$ for a U(1) gauge theory,\\
\beq
\label{bg}
 S_A=-\frac{1}{4}\int \sqrt{-G} F^{MN}F_{MN} 
\eeq
where $F_{MN}$ is the five-dimensional field strength tensor given by\\
\beq
F_{MN}=\partial_M A_N-\partial_N A_M
\eeq
and G is the determinant of the five-dimensional metric.
Now, choosing $A_4=0$,
we decompose  $A_{\mu}$ into it's KK mode as,
\beq
A_{\mu}(x,y)=\sum_{n=0}^{\infty} A_{\mu}^n(x)\chi_n(y)
\eeq
Integrating  
the eqn (\ref{bg}) by parts 
the action becomes, 
\beq
\label{4d}
S_A=\int d^4x \sum_{n=0}^{\infty}\sqrt{-g}[-\frac{1}{4}g^{\mu \kappa}g^{\nu \lambda}F_{\kappa \lambda}^n F_{\mu \nu}^n-\frac{1}{2}m_n^2 A_{\lambda}^n A_{\nu}^n]
\eeq
where 
\beq
F_{\mu\nu}^n=\partial_{\mu}A_{\nu}^n-\partial_{\nu}A_{\mu}^n
\eeq
Here $g$ is the determinant of 4-dimensional space-time.\\ In order to obtain the expression (\ref{4d} )
we  require that the y-dependent wave-function should satisfy the orthonormality condition,
\beq
\int_0^\pi dy \chi^m(y)\chi^n(y)=\delta^{mn}
\eeq
along with the differential equation,
\beq
-\frac{d}{dy}(e^{-2 A(y)}\frac{d \chi^n}{dy})=m_n^2 \chi^n
\eeq
The expression in eq.(\ref{4d})describes the  4-dimensional effective action for 
the gauge fields $A_{\mu}$ with  KK mode masses $m_n$.\\
Transforming the variable $z_n=\frac{m_n}{k} e^{A}$ we get ,
\beq
\frac{d^2 \chi^n}{dy^2}=k^2 \tanh^2(\ln(\omega/c_1)+k y)[z_n^2 \frac{d^2 \chi^n}{dz_n^2}+z_n \frac{d \chi^n}{dz_n} -z_n  {\cosech}^2(\ln(\omega/c_1)+k y)\frac{d \chi^n}{dz_n}]
\eeq
Now for small $\omega^2$ ,the third term in the right hand side of the 
above equation is negligibly small with respect to the second term and the resulting 
differential equation becomes, 
\beq
\left [z_n^2\frac{d^2}{dz_n^2}-z_n \frac{d}{dz_n} +z_n^2 {\coth}^2(\ln(\omega/c_1)+k y)\right]\chi^n=0
\eeq
%Now, $\frac{m_n^2 }{\omega^2 k^2}\rm{sech}^2(\ln(c_2/\omega)-k y)$ $\sim$ $z_n^2-\frac{k^2 \omega^2}{ m_n^2}z_n^4$ 
\\Again, redefining the variable $f=e^{-A(y)}\chi(z)$, the above  differential eqn. takes the form, 
\beq
\label{11}
\left[z_n^2\frac{d^2}{dz_n^2}+z_n \frac{d}{dz_n} +(z_n^2 -1 +\frac{k^2 \omega^2}{m_n^2}z_n^4)\right]f^n=0
\eeq
where we have neglected terms proportional to $\omega^4$.
Since, the term $\frac{k^2 \omega^2}{m_n^2}z_n^4$ is small compared to $(z_n^2-1)$ , we can treat this term as a small  perturbation . 
The solution of above differential equation turns out to be,
\beq
f^n=\frac{1}{N_n}\left[J_1(z_n)+\alpha_n Y_1(z_n)+\delta(z_n)\right]
\eeq
Here, in absence of the last term, the equation as expected is a first order Bessel equation,~\cite{davou}
and the $\delta(z_n)$ is the contribution due to the perturbation. 
This term has $\omega^2$ dependence so that in the limit $\omega^2 \rightarrow 0$, we get back the flat space unperturbed  solution \cite{davou}.
$\frac{1}{N_n}$ is 
Because of the smallness of the perturbation term the normalization constant 
is taken to be same as in ~\cite{davou}. The solution of eqn.(\ref{11}) therefore can be 
written as
\beq
\chi^n(y)=e^{A}\frac{1}{N_n}\left[J_1(z_n)+\alpha_n Y_1(z_n)+\delta(z_n) \right]
\eeq
where as stated above, $J_1((z_n)$ and $ Y_1(z_n)$ are first order Bessel and Neumann functions and $\alpha_n$ are constant coefficients.
It may be recalled  that for the unperturbed case, the solution is
\beq
\chi^n(y)=e^{A}\frac{1}{N_n}[J_1(z_n)+\alpha_n Y_1(z_n)]
\eeq
Hermiticity of the differential operator in above eqn. requires that $\chi_n$ and it's  first derivative should be 
continuous at the orbifolded fixed 
points viz. $y=0$ and $y=\pi$.This leads to, 
\beq
\alpha_n=-\frac{\pi}{2[\ln(\frac{x_n}{2}-A_{\pi}+\gamma+\frac{1}{2}]}
\eeq
where $x_n=\frac{m_n}{k}e^{A_\pi}$ and $A_\pi$ is the value of the warp factor at $y=\pi$ which is $10^{-16}$. 
Here because of the correction to $\alpha_n$ beyond $\omega^2$ has been neglected.\\
Now , we have the differential eqn.for $\delta(z_n)$ as (substituting $f^n$ in eqn.(\ref{11}))
\beq
z_n^2\frac{d^2 \delta}{dz_n^2}+z_n \frac{d \delta}{dz_n}+(z_n^2 -1)\delta+(\frac{k^2 \omega^2}{m_n^2}z_n^4)(J_1(z_n)+\alpha_n Y_1(z_n))=0
\eeq
 Taking the leading order terms in $J_1(z_n)$ and $\alpha_n Y_1(z_n)$ , the equation turns out to be 
\beq
\left[z_n^2\frac{d^2}{dz_n^2}+z_n \frac{d}{dz_n}+(z_n^2 -1)\right]\delta(z_n)+\frac{k^2 \omega^2}{m_n^2}z_n^5=0
\eeq
The  solution of the above equation can be written as,
\bea
\delta(z_n)&=&\frac{1}{2}[(-4 a^2 \pi z_n^3 J_3(z_n) Y_1(z_n)+a^2 \pi z_n^4 J_4(z_n)Y_1(z_n) \\ \nonumber
&+&16 a^2 \pi J_1(z_n)MeigerG[\{\{1\},\{\frac{3}{2}\}\},\{\{2,3\},\{0,\frac{3}{2}\}\},z_n/2,1/2]]
\eea
The perturbed solution is therefore,
\beq
\chi^n(y)=\frac{e^{A}[J_1(z_n)+\alpha_n Y_1(z_n)+\delta(z_n)]}{N_n}
\eeq
Now, we calculate $\frac{d \chi}{dy}$ at $y=\pi$ and put it equal to zero. The roots of this equation 
yield the modified  values of the KK modes 
in the generalized RS scenario. Calculating the derivative at $y=\pi$ ,we get
\beq
x_n^N[J_1'(x_n^N)+\alpha_n Y_1'(x_n^N)+\delta'(x_n^N)]+[
J_1(x_n^N)+\alpha_n Y_1(x_n^N)+\delta(x_n^N)]=0
\eeq
where $(x_n^N)$ denotes the root for the perturbed case. Now, expanding 
\beq 
J_1'(x_n^N)=J_1'(x_n^0)+\partial_{x_n}J_1'(x_n)|_{x_n^0}\frac{ e^A}{k} \Delta{m_n}
\eeq
we obtain  $\Delta{m_n}$ which in turn  gives the value of the shifted root where $x_n^0$ are  the  roots for  
unperturbed solution \cite{davou}. Using the similar Taylor series expansion,
for $\delta(x_n^N)=\delta(x_n^0)$  (Keeping only up to the leading order term ), we finally  arrive at,
\beq
\Delta{m_n}=\frac{x_n^0 \delta'(x_n^0)e^{-A_\pi}k+\delta(x_n^0)e^{-A_\pi}k}{x_n^0 J_1''(x_n^0)+x_n^0 \alpha_n Y_1''(x_n^0)+2 J_1'(x_n^0)+2 \alpha_n Y_1'(x_n^0)}
\eeq
So, corresponding to the old roots $x_1^0=2.45,x_2^0=5.57,x_3^0=8.70$, we get the new roots
such that  $\Delta{m_1}=0.0027\star10^{35} \omega^2$;
$\Delta{m_2}=0.0108\star10^{35} \omega^2$;$\Delta{m_3}=0.0144\star10^{35} \omega^2$ etc.\\
Therefore, $\Delta{m_n}$ gives the correction to the unperturbed kk mode masses for the bulk gauge field in the generalized RS scenario.

\subsection{KK Modes for the anti-symmetric two form Kalb Ramond field}
Here we consider rank-2 antisymmetric tensor field (Kalb-Ramond field ) 
together with gravity in the bulk. It is well known that the third rank tensor field strength corresponding to the 
Kalb-Ramond field can be identified with space-time torsion\cite{ssg}.
The rank-3 antisymmetric field strength tensor $H_{MNL}$  is related to the KR field $B_{MN}$\cite{kalb} as
\beq
\label{1}
H_{MNL} = \partial_{[M}B_{NL]}
\eeq
The 5-dimensional action for the curvature-torsion sector in this case is
\bea
{\cal S}_G=\int d^4x\int dy~\sqrt{-G}~2~M^3~ R(G,H)
\eea
Now, this action can be decomposed into two independent parts -- one consisting of pure curvature and the other, of torsion,
\bea
{\cal S}_G=\int d^4x\int dy \sqrt{-G}~2[M^3~ R(G)-H_{MNL}~H^{MNL}]
\eea
with $H_{MNL}$ related to KR field $B_{NL}$ as in (\ref{1}).
The 5-dimensional action corresponding to the KR field therefore is given by
\beq
{\cal S}_H=\int d^4x \int dy \sqrt{-G}H_{MNL}H^{MNL}
\eeq
As we use the gauge fixing condition in case of bulk gauge field, here also we use KR gauge
fixing condition $B_{4 \mu}=0$.
We now explicitly use the generalized RS metric to calculate the 
above action with the KK mode decomposition for the Kalb-Ramond field as,
\beq
B_{\mu\nu}(x,y)=\sum_{n=0}^{\infty} B_{\mu\nu}^n(x)\chi^n(y)
\eeq
In terms of the 4-D projections $B_{\mu\nu}^n$, an effective action of the form 
\bea
{\cal S}_H=\int d^4x~\sum_{n=0}^{\infty}~
\sqrt{-g}[g^{\mu\alpha}g^{\nu\beta}
g^{\lambda\gamma}H^n_{\mu\nu\lambda}H^n_{\alpha\beta\gamma}+3m_n^2g^{\mu\alpha}g^{\nu\beta}B^n_{\mu\nu}B^n_{\alpha\beta}]
\eea
can be obtained provided the following equation  is satisfied,
\bea
\label{2}
-\frac{d^2\chi^n}{dy^2}=m_n^2\chi^n e^{2 A(y)}
\eea
along with  the following orthogonality condition,
\bea
\int e^{2A(y)} \chi^m(y)\chi^n(y)dy=\delta_{mn}
\eea
Here $H^n_{\mu\nu\lambda}~=~\partial_{[\mu}B^n_{\nu\lambda ]}$ and 
$\sqrt{3}m_n$ gives the mass of the $n$th mode.
In terms of $z_n~=~{\frac{m_n}{k}}e^{A(y))}$, equation (\ref{2}) 
can be recast in the form 
\beq
\left[z_n^2\frac{d^2}{dz_n^2}+z_n\frac{d}{dz_n}-\frac{a^2 z_n^3}{(1-a^2 z_n^2)}\frac{d}{dz_n}+\frac{z^2}{1-a^2 z_n^2}\right]\chi^n=0
\eeq
where $a^2=\frac{k^2 \omega^2}{m
_n^2}$.
Applying the same argument (as is done for the bulk gauge field), the 3rd term is small compared to $z_n$ , hence 
this term is neglected. Treating the last term perturbatively, we arrive at the following differential equation,
\beq
\label{3}
\left[z_n^2\frac{d^2}{dz_n^2}+z_n\frac{d}{dz_n}+z_n^2(1+a^2 z_n^2)\right]\chi^n=0
\eeq
The soln. of the above equation can be written as,
\beq
\label{4}
\chi^n(y)=1/N_n\left[J_0(z_n)+\alpha_n Y_0(z_n)+\delta(z_n)\right]
\eeq
The normalization constant can be found out from the orthogonality condition. 
$\delta(z_n)$ is the perturbation term and it contains $\omega^2$ factor, so that as $\omega^2 \rightarrow 0$ we get back the 
unperturbed solution ~\cite{ssg,ssg1}.
Now, from the continuity conditions at $y=0$ and  $y=\pi$, $\alpha_n$ and $m_n$ can be found out. Using the fact 
that $e^{-A_\pi}>>1$  and  $m_n<<k$, we obtain from the continuity condition at $y=0$,
\beq
\alpha_n \simeq x_n e^{-A_\pi}
\eeq
with $x_n=z_n(\pi)$.
As $\omega^2$ is small, it's dependence on $\alpha_n$ can be neglected ( just as in case of bulk gauge field ). 
Again since, $x_n \simeq 1$ , $\alpha_n$ can be neglected ($\alpha_n<<1$).\\ 
We recall that for the unperturbed case, the boundary 
condition at $y=\pi$ gives, 

\begin{equation}
J_1(x_n)\simeq \frac{\pi}{2}x_n e^{-A_\pi} 
\end{equation}
As the right-hand side of the above equation is negligibly small ,the roots can be approximated to the zeroes of $J_1(x_n)$.

Now, keeping this approach in mind, we consider the perturbed case. 
We are interested  to find how the roots are modified if $J_1(x_n^N)=0$ where $x_n^N$ are the new  roots because of the
modified warp factor in the generalized RS case.\\
If we substitute the solution (\ref{4}) in the eqn.(\ref{3}) we get the differential  for $\delta(z_n)$ as
\beq
\left[z_n^2\frac{d^2}{dz_n^2}+z_n \frac{d}{dz_n}+z_n^2 \right]\delta(z_n) + a^2 z_n^4 J_0(z_n)=0
\eeq
now, to the leading order, $J_0(z_n)= \frac{1}{2}$ and  the above  differential equation can be rewritten  as,
\beq
\left[z_n^2\frac{d^2}{dz_n^2}+z_n \frac{d}{dz_n}+z_n^2 \right]\delta(z_n)+ \frac{1}{2}a^2 z_n^4=0
\eeq
The solution of the above differential equation is given by,
\bea
\delta(z_n)&=&\frac{1}{4}[2 a^2 \pi z_n^2 Y_0(z_n)-4 a^2 \pi z_n J_0(z_n)J_1(z_n)Y_0(z_n)-2 a^2 \pi z_n^2 J_2(z_n)Y_0(z_n) \\ \nonumber
&+& a^2 \pi z_n^3 J_3(z_n)Y_0(z_n)-4 a^2 \pi z_n J_0(z_n)Y_1(z_n)+a^2 \pi z_n^3 J_0(z_n)Y_1(z_n)+4 a^2 \pi z_n J_0(z_n)^2 Y_1(z_n)]
\eea 

Since, $\alpha_n$ is negligibly small, we take the solution as
\beq
\chi^n(y)=J_0(z_n)+\delta(z_n)
\eeq
Performing the derivative at $y=\pi$, we get
\beq
\chi'^n(y)=-J_1'(x_n)+\delta'(x_n)
\eeq
Requiring the fact that $\chi^n(y)$ is zero, we get
\beq
J_1'(x_n^N)+\delta'(x_n^N)=0
\eeq.
Performing the Taylor series expansion w.r.t $x_n^0$ (as was done in case of the  
bulk gauge field ) we find the mass correction as
\beq
\Delta{m_n}=\frac{ \delta'(x_n^0)e^{-A_\pi}k}{ J_2(x_n^0)}
\eeq
$\Delta m_1=0.000647 \star 10^{35}\omega^2$;$\Delta m_2=0.000475 \star 10^{35}\omega^2$;$\Delta m_3=0.000395 \star 10^{35}\omega^2 $ 
corresponding to $m_1=3.83,m_2=7.015,m_3=10.173$ respectively.Here $m_1, m_2,m_3$ denotes the masses 
for the unperturbed case. Thus $\Delta m_n$ gives the correction to the unperturbed kk masses in case of 
Kalb-Ramond field in the generalized RS scenario. 
\subsection{KK modes for the Graviton Field}
Here, same approach is taken as before. Parametrization of the 
tensor fluctuations $h_{\alpha \beta}$ has been done by taking a linear expansion of the flat metric about it's Ricci flat 
value,$\hat{G}_{\alpha \beta}=e^{-2 A(y)}(g_{\alpha \beta}+\kappa^* h{\alpha \beta})$,
where $\kappa^* $ is an expansion parameter~\cite{davo,davo1} In order to calculate mass spectrum of tensor fluctuations, 
we consider 4-dimensional $\alpha \beta$ components with the replacement $G_{\alpha \beta}\rightarrow \hat{G}_{\alpha \beta}$, 
keeping terms up to $\bigcirc(\kappa^*)$. We work in the gauge with $h_{\alpha}^{\alpha}=0$. Expanding  $h_{\alpha \beta}$  
into a KK tower
\beq
h_{\alpha \beta}(x,y)=\sum_{n=0}^{\infty}h_{\alpha \beta}^n(x)\chi^n(y)
\eeq.
we obtain the equation of motion of $h_{\alpha \beta}^n$ as,
\beq
\sqrt{-g}(g^{\alpha \beta}\partial_{\alpha}\partial_{\beta}-m_n^2)h_{\mu\nu}^n(x)=0
\eeq
In order to obtain the eqn. in the above form, the following two conditions must be satisfied for $\chi^n(y)$
\beq
-\frac{d}{dy}(e^{-4 A(y)}\frac{d \chi^n}{dy})=m_n^2 e^{-2 A(y)} \chi^n
\eeq
and 
\begin{equation}
\int e^{-2A(y)} \chi^m(y)\chi^n(y)dy=\delta_{mn}
\end{equation}
The latter defines the orthogonality relation for $ \chi^m(y) $.
Now transforming the variable $z_n~=~{\frac{m_n}{k}}e^{A(y))}$ and defining $\chi(z)=e^{2 A}f$ 
we get the differential equation as, 
\beq
\left[z_n^2\frac{d^2 \delta}{dz_n^2}+z \frac{d \delta}{dz_n}+(z_n^2 -4)\delta+\frac{k^2 \omega^2}{m_n^2}z_n^4\right]f^n=0
\eeq
Since, as stated before, $\frac{k^2 \omega^2}{m_n^2}$ is very small, we can write the solution of the above differential equation as
\beq
f^n=\frac{1}{N_n}\left[J_2(z_n)+\alpha_n Y_2(z_n)+\delta(z_n)\right]
\eeq
Since for the unperturbed case~\cite{davo} we had $\alpha_n Y_2(z_n)$ to be very small in the limit $m_n/k<<1$ and $e^{A_\pi}>>1$, here 
also we neglect that dependence and write the solution as
\beq
f^n=\frac{1}{N_n}\left[J_2(z_n)+\delta(z_n)\right]
\eeq
Putting this solution into the above equation, we obtain the  equation  as,
\beq
\left[z_n^2\frac{d^2}{dz^2}+z_n \frac{d}{dz}+(z_n^2 -4)\right]\delta+\frac{1}{8}\frac{k^2 \omega^2}{m_n^2}z_n^6=0
\eeq
as $J_2(z_n)=\frac{1}{8}z_n^2$.(to the leading order)
The solution of $\delta(z_n)$ is given by,
\bea
\delta(z_n)&=&\frac{1}{16}[(-6 a^2 \pi z_n^4 J_4(z_n) Y_2(z_n)+a^2 \pi z_n^5 J_5(z_n)Y_2(z_n) \\ \nonumber
&+& 32 a^2 \pi J_2(z_n)MeigerG[\{\{1\},\{\frac{3}{2}\}\},\{\{2,4\},\{0,\frac{3}{2}\}\},z_n/2,1/2]]
\eea

The complete  solution therefore becomes
\beq
\chi^n(y)=\frac{e^{2A}\left[J_2(z_n)+\delta(z_n) \right]}{N_n}
\eeq
Now, just as we have done in the gauge field case, We  calculate the derivative of $\chi(y)$ with respect 
to $y$ at $y=\pi$ and set it to zero. 
From the shifted root, we  get the correction to the different KK modes.\\
Calculating the derivative at $y=\pi$ we get
\beq
x_n^N[J_2'(x_n^N)+\delta'(x_n^N)]+2
[J_2(x_n^N)+\delta(x_n^N)]=0
\eeq
where $(x_n^N)$ denotes the new root. Since we can write,
\beq 
J_2'(x_n^N)=J_2'(x_n^0)+\partial_{x_n}J_2'(x_n)|_{x_n^0}\frac{ e^A}{k} \Delta{m_n}
\eeq
and 
\beq
x_n^N=x_n^N-x_n^0+x_n^0=\Delta{m_n}\frac{e^A}{k}+x_n^0
\eeq
Where $x_n^0$ are the  roots obtained in the unperturbed case. Treating $\delta(x_n^N)=\delta(x_n^0)$, we arrive at, 
\beq
\Delta{m_n}=\frac{x_n^0 \delta'(x_n^0)e^{-A_\pi}k+2 \delta(x_n^0)e^{-A_\pi}k}{x_n^0 J_2''(x_n^0) +3 J_2'(x_n^0)}
\eeq

From these we obtain the correction to the mass 
values $\Delta m_1=0.02105 \star 10^{35}\omega^2$;$\Delta m_2=0.11751 \star 10^{35}\omega^2$;$\Delta m_3=.30584 \star 10^{35}\omega^2$ 
corresponding to $m_1=3.83,m_2=7.015,m_3=10.173$ respectively which are  the roots of $J_1(x_n^0)$.

\begin{table}[h]
\renewcommand{\tabcolsep}{0.5pc} % enlarge column spacing
\begin{tabular}{|c|c|c|c|c|c|c|} \hline
Different  & \multicolumn{2}{c|}{$n=1 $} & \multicolumn{2}{c|}{$n=2$} & \multicolumn{2}{c|}{$n=3$} \\ 
\cline{2-7}
bulk fields & $m_n$(in TeV) &  $\Delta{m_n}$(in TeV)  &  $m_n$(in TeV)  &  $\Delta{m_n}$(in TeV)  &  $m_n$(in TeV)  &  $\Delta{m_n}$(in TeV)   \\ \hline
   Graviton field & 3.83 &0.021$\times 10^{35}\omega^2$ &7.02 & 0.117$\times 10^{35}\omega^2$ 
& 10.17 
& 0.305$\times 10^{35}\omega^2$  \\ \hline
   Kalb-Ramond & 6.63 & 0.00064$\times 10^{35}\omega^2$ & 12.14 & 0.00047$\times 
10^{35}\omega^2$ & 17.28 & 
0.00039$\times 10^{35}\omega^2$ \\ \hline
  Gauge -field  & 2.45 &0.0027$\times 10^{35}\omega^2$ & 5.57 & 0.0108$\times 10^{35}\omega^2$ 
& 8.7 & 
0.0144$\times 10^{35}\omega^2$ \\ \hline
\end{tabular}
\caption{ The masses of a few low-lying  modes for different tensorial fields for $kr_c=12$ and
 $k=10^{19}$GeV in case of Ads space. } \label{table1}
\end{table}

\section{KK modes for different bulk fields for de-Sitter brane ($\Omega >0$)}
We can calculate the similar mass splittings in this generalized scenario when the  
induced brane cosmological constant $\Omega >0$ . The warp factor in this case is 
\beq
 e^{-A} = \omega \sinh\left(\ln \frac {c_2}{\omega} - ky \right)~,
\eeq  
where $c_2=1+\sqrt{1+\omega^2}$.Due to different warp factor , the differential equation  for the perturbed solution
$\delta(z_n)$ changes and thereby it's solution will change automatically.\\
\subsection{KK modes for the bulk Gauge field}
Here Applying the exact procedure as in case of ADS bulk, we get the differential equation as
\beq
\left[z_n^2\frac{d^2}{dz^2}+z_n \frac{d}{dz}+(z_n^2 -1)\right]\delta-\frac{k^2 \omega^2}{m_n^2}z_n^5=0
\eeq
The  solution of the above differential equation as,
\bea
\delta(z_n)&=&\frac{1}{2}[(4 a^2 \pi z_n^3 J_3(z_n) Y_1(z_n)-a^2 \pi z_n^4 J_4(z_n)Y_1(z_n) \\ \nonumber
&-&16 a^2 \pi J_1(z_n)MeigerG[\{\{1\},\{\frac{3}{2}\}\},\{\{2,3\},\{0,\frac{3}{2}\}\},z_n/2,1/2]]
\eea

The correction to the mass  is given by the equation 
\beq
\Delta{m_n}=\frac{x_n^0 \delta'(x_n^0)e^{-A_\pi}k+\delta(x_n^0)e^{-A_\pi}k}{x_n^0 J_1''(x_n^0)+x_n^0 \alpha_nY_1''(x_n^0)+2 J_1'(x_n^0)+2 \alpha_n(x_n^0)}
\eeq
So, corresponding to the unperturbed  roots $x_1^0=2.45,x_2^0=5.57,x_3^0=8.70$, we get the new roots
at $\Delta{m_1}=0.0027\star10^{35} \omega^2$;
$\Delta{m_2}=0.0108\star10^{35} \omega^2$;$\Delta{m_3}=0.0144\star10^{35} \omega^2$ etc.\\

\subsection{KK modes for the Kalb-Ramond field}
In this case, The differential equation will be given by
\beq
\left[z_n^2\frac{d^2}{dz_n^2}+z_n \frac{d}{dz_n}+z_n^2 \right]\delta(z_n)- a^2 z_n^4 J_0(z_n)=0
\eeq
So the solution is now given by,
\bea
\delta(z_n)&=&\frac{1}{4}[-2 a^2 \pi z_n^2 Y_0(z_n)+4 a^2 \pi z_n J_0(z_n)J_1(z_n)Y_0(z_n)+ 2 a^2 \pi z_n^2 J_2(z_n)Y_0(z_n) \\ \nonumber
&-& a^2 \pi z_n^3 J_3(z_n)Y_0(z_n)+4 a^2 \pi z_n J_0(z_n)Y_1(z_n)-a^2 \pi z_n^3 J_0(z_n)Y_1(z_n)-4 a^2 \pi z_n J_0(z_n)^2 Y_1(z_n)]
\eea

applying the same procedure as before, we get $\Delta m_n$ as
\beq
\Delta m_n=\frac{\delta'(x_n^0)}{J_2(x_n^0)}k e^{-A_\pi}
\eeq.
So, for same values of $x_1^0,x_2^0,x_3^0$(which has been written in case of ADS bulk, we get
$\Delta m_1=0.000647 \star 10^{35}\omega^2$;$\Delta m_2=0.000475 \star 10^{35}\omega^2$;$\Delta m_3=0.000395 \star 10^{35}\omega^2 $ \\

\subsection{KK modes for the Graviton field}
The last case that we have studied is the Graviton in the ADS bulk. In this case, the differential equation is given by
\beq
\left[z_n^2\frac{d^2}{dz^2}+z_n \frac{d}{dz}+(z_n^2 -4)\right]\delta-\frac{1}{8}\frac{k^2 \omega^2}{m_n^2}z_n^6=0
\eeq
as $J_(z_n)=\frac{1}{8}z_n^2$(to the leading order),
The solution of $\delta(z_n)$ is given by,
\bea
 \delta(z_n)&=&\frac{1}{16}[(6 a^2 \pi z_n^4 J_4(z_n) Y_2(z_n)-a^2 \pi z_n^5 J_5(z_n)Y_2(z_n) 
\\ \nonumber
&-& 32 a^2 \pi J_2(z_n)MeigerG[\{\{1\},\{\frac{3}{2}\}\},\{\{2,4\},\{0,\frac{3}{2}\}\},z_n/2,1/2]]
\eea

$\Delta m_n$ in this case, is given by,
\beq
\Delta{m_n}=\frac{x_n^0 \delta'(x_n^0)e^{-A_\pi}k+2 \delta(x_n^0)e^{-A_\pi}k}{x_n^0 J_2''(x_n^0) +3 J_2'(x_n^0)}
\eeq
from these we get the correction to the mass values $\Delta m_1=0.02105 \star 10^{35}\omega^2$;$\Delta m_2=0.11751 \star 10^{35}\omega^2$;$\Delta m_3=0.30584 \star 10^{35}\omega^2$ corresponding to $m_1=3.83,m_2=7.015,m_3=10.173$ respectively which are basically the roots of $J_(x_n^0)$.

\begin{table}[h]
\renewcommand{\tabcolsep}{0.5pc} % enlarge column spacing
\begin{tabular}{|c|c|c|c|c|c|c|} \hline
Different  & \multicolumn{2}{c|}{$n=1 $} & \multicolumn{2}{c|}{$n=2$} & \multicolumn{2}{c|}{$n=3$} \\ 
\cline{2-7}
bulk fields & $m_n$(in TeV) &  $\Delta{m_n}$(in TeV)  &  $m_n$(in TeV)  &  $\Delta{m_n}$(in TeV)  &  $m_n$(in TeV)  &  $\Delta{m_n}$(in TeV)   \\ \hline
   Graviton field & 3.83 &0.021$\times 10^{35}\omega^2$ &7.02 & 0.117$\times 10^{35}\omega^2$ 
& 10.17 
& 0.305$\times 10^{35}\omega^2$  \\ \hline
   Kalb-Ramond & 6.63 & 0.0006$\times 10^{35}\omega^2$ & 12.14 & 0.00047$\times 
10^{35}\omega^2$ & 17.28 & 
0.00039$\times 10^{35}\omega^2$ \\ \hline
  Gauge -field  & 2.45 &0.0027$\times 10^{35}\omega^2$ & 5.57 & 0.0108$\times 10^{35}\omega^2$ 
& 8.7 & 
0.0144$\times 10^{35}\omega^2$ \\ \hline
\end{tabular}
\caption{ The masses of a few low-lying  modes for different tensorial fields for $kr_c=12$ and
 $k=10^{19}$GeV in case of ds space. } \label{table2}
\end{table}

\section{Conclusion}
In this work we have derived the modifications of the KK mode masses for the various bulk fields because of
a non-zero cosmological constant ( $\omega^2$ ) on the visible brane. We have shown that for both negative and positive brane 
cosmological constant ( i.e for 
anti de-Sitter and de-Sitter universe ) the leading order corrections to the KK mode masses are proportional to $\omega^2$.
It was discussed earlier that while for de-Sitter case there is no bound for the value of the cosmological constant, for
anti de-Sitter case the magnitude can be at most $\sim 10^{-32}$ if one wants to resolve the hierarchy problem
consistently. Though the present observed value of the cosmological constant is tiny and positive with $\omega^2 \sim 10^{-124}$ ( in Planck
unit ), it may however be argued that an anti de-Sitter universe with a relatively large negative brane cosmological constant
say $\sim  10^{-32}$ ( inherited from the bulk ) 
may subsequently have evolved into a de-Sitter universe with  a tiny positive value of the brane cosmological constant because of other
effects on the brane which may induce positive cosmological constant on the brane. Such a scenario
will then indicate the possibility of a large correction to the KK modes of the bulk fields resulting into 
significant modifications ( from a flat brane Randall-Sundrum scenario) of experimental signatures
of various processes involving the bulk and the standard model fields.

\acknowledgements{JM acknowledges CSIR, Govt. of India for providing 
financial support. JM would like to thank S. Majhi for useful suggestions.}


\begin{thebibliography}{99}
\bibitem{ADD} N. Arkani-Hamed, S. Dimopoulos and G. Dvali, 
Phys. Lett. {\bf B 429}, 263 (1998); {\em ibid} Phys.Rev. {\bf D 59}, 086004 (1999);
I. Antoniadis, N. Arkani-Hamed, S.
Dimopoulos and G. Dvali, Phys. Lett. {\bf B436} 257 (1998).K.Akama,
Prog. Theor. Phys. 80, 935 (1988)


\bibitem{cohen} A. G. Cohen and D. B. Kaplan, Phys. Lett. {\bf
B470} , 52 (1999); I. Antoniadis, S. Dimop oulos and A. Giveon,
JHEP, {\bf 05}, 055 (2001); T. Multamaki and  I. Vilja, Phys.
Lett. {\bf B545}, 389 (2002); C. P. Burgess, J. M. Cline, N. R.
Constable and H. Firouzjahi, JHEP, {\bf 01}, 014 (2002)

\bibitem{RS} L. Randall and R. Sundrum, Phys. Rev. Lett. {\bf 83}, 3370 (1999)


\bibitem{GW} W.~D.~Goldberger and M.~B.~Wise,
  %``Modulus stabilization with bulk fields,''
  Phys.\ Rev.\ Lett.\  {\bf 83}, 4922 (1999)


\bibitem{gwpheno} W.~D.~Goldberger and M.~B.~Wise,
  %``Bulk fields in the Randall-Sundrum compactification scenario,''
  Phys.\ Rev.\  D {\bf 60}, 107505 (1999)


\bibitem{davou}H.~Davoudiasl, J.~L.~Hewett and T.~G.~Rizzo,
  %``Bulk gauge fields in the Randall-Sundrum model,''
  Phys.\ Lett.\  B {\bf 473}, 43 (2000)


\bibitem{davou1}H.~Davoudiasl, J.~L.~Hewett and T.~G.~Rizzo,
  %``Brane localized kinetic terms in the Randall-Sundrum model,''
  Phys.\ Rev.\  D {\bf 68}, 045002 (2003)


\bibitem{davou2}H.~Davoudiasl, J.~L.~Hewett and T.~G.~Rizzo,
 %``Brane localized curvature for warped gravitons,''
  JHEP {\bf 0308}, 034 (2003)


\bibitem{agashe} K.~Agashe, N.~G.~Deshpande and G.~H.~Wu,
  %``Can extra dimensions accessible to the SM explain the recent measurement of
  %anomalous magnetic moment of the muon?,''
  Phys.\ Lett.\  B {\bf 511}, 85 (2001)
  [arXiv:hep-ph/0103235]

 \bibitem{davo}H.~Davoudiasl, J.~L.~Hewett and T.~G.~Rizzo,
  %``Phenomenology of the Randall-Sundrum gauge hierarchy model,''
  Phys.\ Rev.\ Lett.\  {\bf 84}, 2080 (2000)

\bibitem{gsw}{\em Superstring Theory}, M. B. Green, J. H. Schwarz and 
E. Witten, Cambridge University Press, Cambridge (1987)


\bibitem{lebedev} O.~Lebedev,
  %``Torsion Constraints in the Randall--Sundrum Scenario,''
  Phys.\ Rev.\  D {\bf 65}, 124008 (2002) [arXiv:hep-ph/0201125]

%\bibitem{gopal} K.~Agashe, S.~Gopalakrishna, T.~Han, G.~Y.~Huang and A.~Soni,
  %``LHC Signals for Warped Electroweak Charged Gauge Bosons,''
 % arXiv:0810.1497 [hep-ph]
\bibitem{little} H.~Davoudiasl,
  %``The Little Randall-Sundrum Model at the LHC,''
  arXiv:0810.0194 [hep-ph]

\bibitem{bombi}C.~Bambi and F.~R.~Urban,
  %``Gravitational production of KK states,''
  Phys.\ Rev.\  D {\bf 78}, 103515 (2008)
  [arXiv:0808.3500 [hep-ph]]

\bibitem{gunion} B.~Grzadkowski and J.~F.~Gunion,
  %``KK gravitons and unitarity violation in the Randall-Sundrum model,''
  Phys.\ Lett.\  B {\bf 653}, 307 (2007)
%\bibitem{lhc} K.~Agashe {\it et al.},
  %``LHC Signals for Warped Electroweak Neutral Gauge Bosons,''
 % Phys.\ Rev.\  D {\bf 76}, 115015 (2007)
  %[arXiv:0709.0007 [hep-ph]].
\bibitem{rizzo1} T.~G.~Rizzo,
 %``Black hole production at the LHC by standard model bulk fields in the
  %Randall-Sundrum model,''
  Phys.\ Lett.\  B {\bf 647}, 43 (2007)
  [arXiv:hep-ph/0611224]

\bibitem{liam}A.~L.~Fitzpatrick, J.~Kaplan, L.~Randall and L.~T.~Wang,
  %``Searching for the Kaluza-Klein Graviton in Bulk RS Models,''
  JHEP {\bf 0709}, 013 (2007)
%\bibitem{muta} N.~Muta and N.~Uekusa,
  %``Localized Kaluza-Klein graviton and cosmological constant,''
 % arXiv:hep-ph/0503017
\bibitem{dominici}D.~Dominici, B.~Grzadkowski, J.~F.~Gunion and M.~Toharia,
  %``The scalar sector of the Randall-Sundrum model,''
  Nucl.\ Phys.\  B {\bf 671}, 243 (2003)
  [arXiv:hep-ph/0206192]

\bibitem{hawking} A.~Chamblin, S.~W.~Hawking and  H.~S.~Reall, Phys.Rev.D {\bf 61},065007 (2000)

\bibitem{genrs} S. Das, D. Maity and S. Sengupta, JHEP {\bf 05}, 042 (2008)


\bibitem{christof} C.~Schmidhuber,  %``AdS(5) and the 4d cosmological constant,''
  Nucl.\ Phys.\  B {\bf 580}, 140 (2000)
  [arXiv:hep-th/9912156]

\bibitem{kaloper} N. Kaloper Phys.\ Rev.\  D {\bf 60}, 123506 (1999)

\bibitem{jm} R.~Koley, J.~Mitra and S.~SenGupta,
  %``Fermion localization in generalised Randall Sundrum model,''
  Phys.\ Rev.\  D {\bf 79}, 041902 (2009)
  [arXiv:0806.0455 [hep-th]]


\bibitem{koley} R.~Koley, J.~Mitra and S.~SenGupta,
  %``Modulus stabilization of generalized Randall Sundrum model with bulk scalar
  %field,''
  Europhys.\ Lett.\  {\bf 85}, 41001 (2009)
  [arXiv:0809.4102 [hep-th]]


\bibitem{muta} N.~Muta and N.~Uekusa,
  %``Localized Kaluza-Klein graviton and cosmological constant,''
  arXiv:hep-ph/0503017




\bibitem{param} Y.~Aghababaie, C.~P.~Burgess, S.~L.~Parameswaran and F.~Quevedo,
  %``Towards a naturally small cosmological constant from branes in 6D
  %supergravity,''
  Nucl.\ Phys.\  B {\bf 680}, 389 (2004)
  [arXiv:hep-th/0304256]


%\bibitem{low} K.~Agashe, A.~Falkowski, I.~Low and G.~Servant,
  %``KK Parity in Warped Extra Dimension,''
 % JHEP {\bf 0804}, 027 (2008)
  %[arXiv:0712.2455 [hep-ph]]


\bibitem{maeda} M.~Sasaki, T.~Shiromizu and K.~i.~Maeda,
  %``Gravity, stability and energy conservation on the Randall-Sundrum
  %brane-world,''
  Phys.\ Rev.\  D {\bf 62}, 024008 (2000)


\bibitem{karch} A.~Karch and L.~Randall,
  %``Locally localized gravity,''
  JHEP {\bf 0105}, 008 (2001)


\bibitem{gross}Y.~Grossman and M.~Neubert,
  %``Neutrino masses and mixings in non-factorizable geometry,''
  Phys.\ Lett.\  B {\bf 474}, 361 (2000)


\bibitem{ssg} B.~Mukhopadhyaya, S.~Sen and S.~SenGupta,
  %``Does a Randall-Sundrum scenario create the illusion of a torsion-free
  %universe?,''
  Phys.\ Rev.\ Lett.\  {\bf 89}, 121101 (2002)
  [Erratum-ibid.\  {\bf 89}, 259902 (2002)]


\bibitem{kalb}M.Kalb and P.Ramond, Phys.Rev. D9, 2273 (1974)


\bibitem{ssg1} B.~Mukhopadhyaya, S.~Sen and S.~SenGupta,
  %``Bulk antisymmetric tensor fields in a Randall-Sundrum model,''
  Phys.\ Rev.\  D {\bf 76}, 121501 (2007)


% \bibitem{davo}H.~Davoudiasl, J.~L.~Hewett and T.~G.~Rizzo,
  %``Phenomenology of the Randall-Sundrum gauge hierarchy model,''
 % Phys.\ Rev.\ Lett.\  {\bf 84}, 2080 (2000)


 \bibitem{davo1}  H.~Davoudiasl, J.~L.~Hewett and T.~G.~Rizzo,
  %``Brane localized curvature for warped gravitons,''
  JHEP {\bf 0308}, 034 (2003)


 
\end{thebibliography}
\end{document}